\documentclass[12pt]{article}
    \usepackage{amsmath,amsxtra,amssymb, amsfonts }

\newtheorem{thm}{Theorem}
\newtheorem{cor}[thm]{Corollary}
 
\newtheorem{lemma}[thm]{Lemma}

\newtheorem{prop}[thm]{Proposition}

\newtheorem{exam}{Example}

\def\be{\begin{eqnarray}}
\def\ee{\end{eqnarray}}
\def\bee{\begin{eqnarray*}}
\def\eee{\end{eqnarray*}}
\def\bal{\begin{align}}
\def\enal{\end{align}}
\def\pmx{\begin{pmatrix}}
\def\emx{\end{pmatrix}}
\def\bsq{\begin{subequations}}
\def\esq{\end{subequations}}

\def\ds{\displaystyle}
\def\ts{\textstyle}
\def\nn{\nonumber}

\def\wh{\widehat}
\def\raw{\rightarrow}
 \def\imp{\Rightarrow}

\def\ot{\otimes}

\def\Iff{~{\Longleftrightarrow}~}
\def\tr{{\rm Tr} \, }
\def\trp{{\rm Tr} }
\def\bra{\langle}
\def\ket{\rangle}
\def\kb{ \ket \bra }

\newcommand{\proj}[1]{ | #1 \kb  #1|}
\newcommand{\kt}[1]{ | #1 \ket}
\newcommand{\braa}[1]{ \bra #1 |}
\newcommand{\norm}[1]{ \| #1  \|}

\def\mm{ \! - \!}
\def\pp{ \! + \!}

\def\half{{ \tfrac{1}{2} }}

\def\rt2{\ts \frac{1}{\sqrt{2}} }
\def\rt3{\ts \frac{1}{\sqrt{3}} }

\def\ovb{\overline}

\def\hil{{\cal H}}
\def\id{{\cal I}}

\def\qed{\qquad{\bf QED}}
\def\hv{{\rm Holv}}
\def\prf{\noindent {\bf Proof: }}

\def\bw{{\mathbf w}}

\def\cT{{\mathcal T}}
\def\cP{{\mathcal P}}

\def\cB{{\mathcal B}}

 \def\d2{{d^{\prime}}}
  \def\di{{d_i^{\prime}}}
\def\td{\tfrac{1}{d}}
\def\eof{{\rm EoF}}

\title{Properties of Conjugate Channels with
Applications to Additivity and Multiplicativity }

      \author{Christopher King \\
{\small      Department of Mathematics,
Northeastern University,  Boston MA 02115} \\
   {\small  king@neu.edu} \\
   \and Keiji Matsumoto \\  {\small National Institute of Informatics} \\
{\small  2-1-2 Hitotsubashi, Chiyoda-ku, Tokyo 101-8430, Japan}
 \\ {\small and} \\
{\small ERATO Quantum Computation and Information Project, JST} \\
  {\small  Danai Hongo White Bldg,
    5-28-3 Hongo, Bunkyo-ku, Tokyo 113-0033, Japan} \\
{\small  keiji@nii.ac.jp} \\
   \and  Michael Nathanson \\
{\small      Department of Mathematics, Kenyon College, Gambier, OH 43022} \\
    {\small nathansonm@kenyon.edu} \\
\and Mary Beth Ruskai
        \\  {\small  Department of Mathematics,
     Tufts University,
       Medford, MA 02155} \\
     {\small     Marybeth.Ruskai@tufts.edu}}

\date{ \today \\ ~~ \\ {\em Dedicated to the memory of John  T. Lewis}}

\begin{document}

       \maketitle

\pagebreak

       \begin{abstract}
Quantum channels can be described via a unitary coupling of system and environment,
 followed by a trace over the environment state space. Taking the trace instead over the
 system state space produces a different mapping which we call the conjugate channel. 
 We explore the properties of conjugate channels and describe several different methods
 of construction.    In general, conjugate channels map $M_d \mapsto M_\d2$ with
 $d < \d2$, and different constructions may differ by conjugation with a partial isometry.
   We show that a channel and its conjugate have the same minimal output entropy and 
   maximal output $p$-norm.
It then follows that the additivity and  multiplicativity conjectures for these measures of optimal
output purity hold  for a product of channels if and only if they
 also hold for the product of their conjugates.    This allows us to reduce these
conjectures to the special case of maps taking $M_d \mapsto M_{d^2}$ with a
minimal representation of dimension at most $d$.

We find explicit expressions
for the conjugates for a number
of well-known examples, including entanglement-breaking channels, unital qubit channels,
the depolarizing channel, and a subclass of random unitary channels.
For the entanglement-breaking channels,  channels this yields a new class of channels for 
which  additivity and multiplicativity of optimal output purity can be established.
For  random unitary channels using the generalized Pauli matrices,
we obtain a new formulation of the multiplicativity  conjecture.  The conjugate
of the completely noisy channel plays a special role and suggests a 
mechanism for using noise to transmit information.
           \end{abstract}


\tableofcontents

\section{Introduction}
 
 The underlying model of noise in a quantum system regards the
 original system (typically called Alice) as a subsystem of a larger
 system which includes both the original system and the 
 environment, which we call Bob.    We assume that Alice and Bob initially prepare
 their systems separately after which they evolve in time according to the unitary 
 group  of the Hamiltonian of the combined system.    Either system can be 
 described at a later time by taking a partial trace over the other.    Typically,
 the unitary interaction entangles the two systems so that each subsystem 
 is in a mixed state.       In the most common scenario,
 Alice can prepare a variety of different states, but Bob
 always uses the same state $\proj{\phi}$.    The map  which takes  Alice's
 state $\proj{\psi}$ to $\trp_B  \, U(t) \proj{ \psi \ot \phi} U(t)^\dag \equiv \Phi(\proj{\psi})$ 
 at a fixed time $t$  is called a channel $\Phi$.       Taking
 $\trp_A  \, U(t) \proj{ \psi \ot \phi} U(t)^\dag$ defines a map $\Phi^C$
  whose   output  is a state $\Phi^C (\proj{\psi})$ which describes the
  information available to Bob at the same fixed time $t$.
  We call this map   the {\em conjugate channel.} 
   
 Mathematically, both $\Phi$ and $\Phi^C$ are completely positive 
 trace-preserving (CPT) maps.   The former takes $\cB(\hil_A) \mapsto \cB(\hil_A) $
 and the latter $\cB(\hil_A) \mapsto \cB(\hil_B) $ where  $\hil_A$ and $\hil_B$
 denote the Hilbert spaces of Alice and Bob respectively.
  In this paper we develop the  properties of conjugate channels
for CPT maps when both Hilbert spaces are finite dimensional.
We study  the conjugates of several classes of channels, including  
entanglement-breaking  (EBT) maps
 and a higher-dimensional analog of the qubit unital  channels,
  which we call Pauli-diagonal. 
Conjugate channels have been studied before in other settings including
           Appendix B of \cite{DS} and \cite{W1}.
After completion of this work, we learned that
much of our  analysis of conjugate maps was done independently by
           Holevo    \cite{HvCC} for maps which are completely 
           positive, but not necessarily  trace-preserving.
              Holevo obtains most of the
           results in Sections~\ref{const} and \ref{sect:EBT}, with the
           exception of our Theorem~\ref{thm:ext.red}.  He also   obtains a
           result similar to Corollary~\ref{cor:bob}.

Although the channels  $\Phi$ and $\Phi^C$
are quite different in general, for any pure input state $| \psi \ket$ the two
output states $\Phi(\proj \psi)$ and $\Phi^{C}(\proj \psi)$ must have the same
nonzero spectrum. This means that the channels $\Phi$ and $\Phi^C$ 
have the same
maximal output $p$-norms $\nu_p(\Phi) = \sup_{\rho} || \Phi(\rho) 
||_p = \nu_p(\Phi^{C})$,
and also the same minimal output entropy. We show that for any pair of
  channels, 
the additivity and multiplicativity conjectures for these measure of optimal output
purity hold if and only if the same conjecture holds for their conjugate channels.
For example 
\be
\nu_p(\Phi_1 \ot \Phi_2) = \nu_p(\Phi_1) \nu_p(\Phi_2) \Iff 
 \nu_p(\Phi_1^C \ot \Phi_2^C) = \nu_p(\Phi_1^C) \nu_p(\Phi_2^C).
\ee
This equivalence allows  to obtain some new results about these conjectures.
    One of these is the realization that it would suffice to prove them  
    for the special class of maps taking     $M_d \mapsto M_{d^2}$ with a
minimal representation of dimension at most $d$.

This work was motivated by the observation of a similarity between 
King's proofs of multiplicativity  \cite{King4} for EBT maps and for 
Hadamard diagonal  channels \cite{King5} which seemed to suggest a 
kind of duality.    The concept of conjugate channels allows us to make
this duality explicit when the EBT   map belongs to a subclass we
call extreme CQ and the Hadamard diagonal  channel is also trace-preserving.

A  large part of the paper considers channels which are convex combinations
 of unitary conjugations of the generalized Pauli matrices; we will call
 these channels Pauli diagonal.   We show that
 the image of the conjugate of the completely noisy channel  is essentially
  equivalent to the original state, i.e., when the noise completely destroys Alice's 
  state,   Bob can recover it.    We also show that the conjugate of a
Pauli diagonal channel can be written as the composition of a simple Hadamard
   channel (using only one diagonal Kraus operator) 
  with the conjugate of the completely noisy map.
This allows a simple and  appealing reformulation of the multiplicativity
conjecture for these channels.
 Although it has not yet led to a proof,    this approach 
provides some new insights.

The paper is organized as follows. In Section \ref{const} we define 
the conjugate of a channel, show how to use its Kraus representation 
to construct its conjugate and show that it is well-defined up to
a partial isometry.    We also prove the equivalence of the multiplicativity
problem for a product of channels and their conjugates, and a reduction
theorem.  In Section 
\ref{sect:EBT} we study the conjugates of EBT channels, and extend
King's results \cite{King5} about  Hadamard diagonal channels to 
a larger class, which we call simply Hadamard channels.
 In Section \ref{sect:Pdiag} we study the Pauli diagonal channels and several
 related classes of randomÊ  unitary channels.    In Section~\ref{GL} we
derive a relation between the conjugate channels and the
Giovannetti-Lloyd linearization operators which arose in the study of
multiplicativity for integer values of $p$ \cite{GL}. 
  Appendix~\ref{app:rep} summarizes some
 basic results about representations of channels and extends them
 to the less familiar situations of maps between spaces of different
 dimension and equivalence via partial isometries.
Appendix~\ref{app:qbit}  gives a detailed analysis of the issues
which arise  for  the Pauli diagonal channels  in the case of unital
qubit maps.

\section{Basic definitions and results}\label{const}

\subsection{Construction of conjugate channels}

We begin with two well-known representations of a CPT map 
$\Phi: M_d \mapsto M_{d^\prime}$.   The Lindblad-Stinespring 
(LS) ancilla representation  \cite{Stine,Lind75}  can be written as
\be  \label{LSform}
\Phi(\rho) = \trp_C \, U \, \big( \rho \ot \proj{\phi} \big) \, U^\dag
\ee
where $|\phi \ket$ is a pure state on an auxiliary space $\hil_C$,
and $U : {\Bbb C}^{d} \ot \hil_C \mapsto
{\Bbb C}^{d^\prime} \ot \hil_C$ is a partial isometry.      
We denote the  minimal dimension for the auxiliary space $\hil_C$ 
as $\kappa$ and note that   $\kappa \leq d d^{\prime}$.   There is
no loss of generality in assuming that  the rank of $U$ is $d_\kappa$,
and we will always assume that $d_C = \dim \hil_C < \infty$.  However,
we will not restrict ourselves to minimal representations.

The standard operator-sum or Kraus-Choi representation of $\Phi$ is \cite{Kraus}
\be\label{OSform}
\Phi(\rho) & = & \sum_{k=1}^{d_C} F_k \, \rho \, F_k^{\dag}
\ee
where the operators satisfy the trace-preserving condition $\sum F_{k}^{\dag} F_k = I$.
As discussed in Appendix~\ref{app:rep}, these two representations
can be connected by the relation 
$F_k = \trp_C \, \Big[ U (I \ot  | \phi \ket \bra e_k | )  \Big]$.   Moreover,
different Kraus representations can be related a by partial
isometry  $W$ of rank $\kappa$ as explained after \eqref{K2}.

Using the  representation \eqref{LSform},
we construct the {\em conjugate channel}  $\Phi^C: M_d \mapsto M_{d_C}$ as
\be \label{def:conj}
\Phi^C(\rho) = \trp_B \, U \, \big( \rho \ot \proj{\phi} \big) \, U^\dag
\ee
where we identify $\hil_B = {\Bbb C}^{d^\prime}$. To see how
(\ref{def:conj}) arises from the Kraus representation (\ref{OSform}),
define
\be\label{def:F}
     {\bf F}(\rho) = \sum_{jk} | e_j \kb e_k|  \, \ot  \, F_j \rho F_k^{\dag}
    \,\, \in \,\,  M_{d_C} \ot M_{d^\prime}
\ee
Then $\Phi(\rho) = \trp_C \,  {\bf F}(\rho) $, and  the
conjugate channel can be written as
\be\label{Kr->conj}
     \Phi^C(\rho) = \trp_B \, {\bf F}(\rho) =
        \sum_{jk} \,  \tr \Big( F_j \rho F_k^{\dag}  \Big) \, | e_j \kb e_k| .
\ee
The channel (\ref{Kr->conj}) can itself be written in a Kraus representation
\be\label{conj-Kr}
\Phi^C(\rho) = \sum_{\mu = 1}^{d^\prime} R_{\mu} \, \rho \, R_{\mu}^{\dag}
\ee
where the Kraus operators satisfy
\be\label{conj-Kr}
(R_{\mu})_{j k} = (F_{j})_{\mu k}
\ee
Two sets of Kraus operators related as in \eqref{conj-Kr} will 
also be called {\em conjugate}.

For a given channel $\Phi$, there are many choices possible for
$\hil_C$, $|\phi \ket$ and $U$ in (\ref{LSform}).    We denote by
${\cal C}(\Phi)$ the collection of all conjugate channels defined
as in (\ref{def:conj}) using all representations of the same
channel $\Phi$.    The following Lemma shows that these different
representations are related by conjugation with a partial isometry.

\begin{lemma}\label{lemma1}
For any pair of elements $\Phi^{C_1}, \Phi^{C_2} \in {\cal C}(\Phi)$
one can find a partial isometry $W $   of rank $\kappa$ 
such that  
\be \label{lem1}
\Phi^{C_1}  =   \Gamma_{W } \circ \Phi^{C_2} \qquad \hbox{and} \qquad
\Phi^{C_2}  =   \Gamma_{W^{\dag}} \circ \Phi^{C_1}
\ee
where $ \Gamma_{W}(\rho) = W \rho W^{\dag}$.
\end{lemma}
\prf Let $\{ F_k \}$  and $\{ G_m \}$ be Kraus representations for $\Phi$ so that 
\be
\Phi(\rho) = \sum_{k} F_k \, \rho \, F_k^{\dag}
= \sum_{m} G_m \, \rho \, G_m^{\dag}
\ee
and the corresponding conjugate representations can be written as
\be
\Phi^{C_1}(\rho) = \sum_{\mu}^{d^\prime} R_{\mu} \, \rho \,
R_{\mu}^{\dag}, \quad
\Phi^{C_2}(\rho) = \sum_{\mu}^{d^\prime} S_{\mu} \, \rho \, S_{\mu}^{\dag}.
\ee
with $R_\mu, S_\mu$ given by \eqref{conj-Kr}.    As explained after
\eqref{K2}, 
there is a partial isometry $W$ of rank $\kappa$ such that 
 $F_k = \sum_{k} w_{km } \, G_m$.  Then \eqref{conj-Kr} implies 
that  $R_\mu = W S_\mu$  so that
 \be
\Phi^{C_1}(\rho)  = \sum_{\mu} R_{\mu} \rho R_{\mu}^\dag =
W \,\Big( \sum_{\mu} S_{\mu} \rho S_{\mu}^\dag \Big) \, W^\dag = W  \Phi^{C_2}(\rho)  W^\dag
\ee
If, in addition, $\{ G_m \}$ is minimal, then $W^\dag W = I_{\kappa}$, and it follows
immediately that  $ \Phi^{C_2}(\rho)  = W^\dag  \Phi^{C_1}(\rho) W$.    If neither
representation is minimal, we can use the fact proved after \eqref{K2} that
$G_m = \sum_{j}  \ovb{w}_{jm} F_j $ which implies  $S_\mu = W^\dag R_\mu $
    \qed

In most of our applications and results, the
particular choice of element in ${\cal C}(\Phi)$ will be irrelevant, and
we will generally speak of ``the'' conjugate
channel $\Phi^C$ with the understanding that it is unique up to
the partial isometry described above. With this understanding
we note that the conjugate of the conjugate is the original channel,
that is $(\Phi^{C})^C = \Phi$, or $\Phi \in {\cal C}(\Phi^C)$.
   
   Another method of representing a channel is via its Choi-Jamiolkowski (CJ)
   matrix  \eqref{CJ} which gives a one-to-one correspondence between
   CP  maps $\Phi: M_d \mapsto M_{d^\prime}$ and positive semi-definite
   matrices on $M_d \ot M_ \d2$.   The subset which satisfies 
   $\trp_B \Gamma_{AB} = \tfrac{1}{d} I_d $ gives the CPT maps.
    The next result gives a relation
   between the CJ matrix of a channel and its conjugate.
   \begin{prop}   \label{CJconj}
   Let $\Phi$ be a CPT map with CJ matrix $\Gamma_{AB} = (I \ot \Phi)(\proj{\phi})$
   as in \eqref{CJ} and let  $\Gamma_{ABC}$ be a purification of $\Gamma_{AB} $.
   Then $\Gamma_{A C} = \trp_B \,  \Gamma_{ABC}$ is the CJ matrix of the
   conjugate channel $\Phi^C$.
   \end{prop}
   The proof, which is given in Appendix~\ref{app:rep}, is a consequence of the
   fact that the eigenvectors of the CJ matrix generate a minimal set of 
   Kraus operators.   Although this approach may seem less constructive, in
   some contexts (see work of Horodecki), channels are naturally defined in terms
   of their CJ matrix or ``state representation''.    Moreover, this approach has
   less ambiguity.  If the standard basis for $\hil_C$ is used and  $\Gamma_{AB}$
   has non-degenerate eigenvalues, it is unique up to a permutation.    More 
   generally, if one labels the eigenvalues of    $\Gamma_{AB}$ in increasing
   (or decreasing) order and labels
   the basis for $\hil_C$ accordingly, then $\Gamma_{A C}$ is unique up to   
   conjugation with a unitary matrix
   of the form $I_A \ot  U_C$ where $U_C$ is a unitary matrix which is block
   diagonal corresponding to the degeneracies of $\Gamma_{AB} $.

   \subsection{Optimal output purity}

Our first result, although straightforward,
is a key ingredient, so we state and prove it
explicitly here. 

\begin{thm}   \label{thm:1}
The output  $\Phi(\proj{\psi})$ of a channel acting on a pure state  has
the same non-zero spectrum as the output  $\Phi^C(\proj{\psi})$
of its conjugate acting on the same pure state.
\end{thm}
\prf   Let $\gamma_{AB} = U \big(\proj{\psi} \ot \proj{\phi} \big)
U^\dag = \proj{\Psi_{AB}}$
with $U, |\phi \ket $ as in \eqref{LSform} and $| \Psi_{AB} \ket = U
\big(  |\psi \ket \ot  |\phi \ket \big)$.
Then $\gamma_{AB}$ is a pure state,
$\Phi(\proj{\psi}) = \gamma_A = \trp_B \, \gamma_{AB}$  and
$\Phi^C(\proj{\psi}) = \gamma_B = \trp_A \, \gamma_{AB}$.   The result
then follows from the well-known fact that the reduced density matrices of
a pure state have the same non-zero spectrum.   \qed

As an immediate corollary, it follows that a channel $\Phi$ and its conjugate
$\Phi^C$ always have the same maximal output purity and minimal output
entropy. Recall that the  maximal output purity is defined for $p \geq 1$ by
\be\label{def:nu-p}
\nu_p(\Phi) \equiv \sup_{\rho}  \norm{ \Phi(\rho)}_p
= \sup_{| \psi \ket} \norm{\Phi(\proj{\psi})}_p
\ee
and the minimal output entropy is
\be\label{def:min-ent}
S_{\min}(\Phi)   \equiv \inf_{\rho} S(\Phi(\rho))
= \inf_{| \psi \ket} S(\Phi(\proj{\psi})),
\ee
where the $\sup$ and $\inf$ are taken over normalized states
$\rho$ and $| \psi \ket$.

\begin{cor}  \label{thm:2}
For any CPT map $\Phi$, $\nu_p(\Phi) = \nu_p(\Phi^C) $ and
   $S_{\min}(\Phi) =  S_{\min}(\Phi^C)$.
\end{cor}

For any pair of conjugate channels $\Phi_{1}^C$ and $\Phi_{2}^C$,
the product $\Phi_{1}^C \ot \Phi_{2}^C$ is again a channel,
and from the definition (\ref{def:conj}) it follows that
$\Phi_{1}^C \ot \Phi_{2}^C \in {\cal C}(\Phi_{1} \ot \Phi_{2})$.
Therefore given any representative $[\Phi_1 \ot \Phi_2]^C$ there
is a partial isometry $W$ such that
\be\label{conj-prod}
[\Phi_1 \ot \Phi_2]^C = \Gamma_{W} \circ [\Phi_1^C \ot \Phi_2^C]
\ee
Combining Corollary \ref{thm:2} and (\ref{conj-prod})
implies the equivalence of the additivity and multiplicativity problems for
channels and their conjugates.
For convenience we restate the result below in Theorem \ref{thm:4}.
   \be  \label{mult}
     \nu_p(\Phi_1 \ot \Phi_2) = \nu_p(\Phi_1)  \, \nu_p(\Phi_2)
   \ee
    \be  \label{Sadd}
    S_{\min}(\Phi_1 \ot \Phi_2) = S_{\min}(\Phi_1)   + S_{\min}(\Phi_2)   
    \ee
\begin{thm} \label{thm:4}
For any pair of channels $\Phi_1, \Phi_2$, and any $p \geq 1$,
\eqref{mult} holds if and only if
\bee
    \nu_p(\Phi^C_1 \ot \Phi^C_2) = \nu_p(\Phi^C_1)  \, \nu_p(\Phi^C_2) ,
    \eee
   and  \eqref{Sadd} holds if and only if
\bee S_{\min}(\Phi^C_1 \ot \Phi^C_2) = S_{\min}(\Phi^C_1)   + S_{\min}(\Phi^C_2)
\eee
\end{thm}

There are also additivity conjectures for the
 entanglement of formation (EoF)  and Holevo capacity $C_\hv(\Phi)$.   
 For a bipartite state $\gamma_{AC}$, define
 \be   \label{eof}
\eof(\gamma_{AC})  =  \inf \Big\{ \sum_j \pi_j S\big[ \trp_C \proj{\psi_j} \big] :
             \sum_j \pi_j   \proj{\psi_j} = \gamma_{AC} \Big\}.
             \ee
 Following \cite{MSW} and \cite{Shor2}, we
 use   \eqref{LSform}   to associate $\gamma_{AC}$ with $\Phi$ 
 and a state $\rho$  (as in the proof of Theorem~\ref{thm:1})
 so that $\Phi(\rho)  = \trp_C \, \gamma_{AC}$. 
Then we define
$  \chi[\Phi(\rho)] =  S[\Phi(\rho)] - \eof(\gamma_{AC})  $ and 
$     C_\hv(\Phi)  = \sup_{\rho}  \chi[\Phi(\rho)]$.
  The  additivity conjecture for Holevo capacity is
    \be  \label{Cadd}
    C_\hv(\Phi_1 \ot \Phi_2) = C_\hv (\Phi_1)   + C_\hv(\Phi_2)   
    \ee
and the superadditivity conjecture for EoF is
\be  \label{eofsup}
   \eof(\gamma_{A_1C_1 A_2 C_2} ) \geq    \eof(\gamma_{A_1C_1} )  
    +    \eof(\gamma_{A_2 C_2} ) .  
    \ee 
   Shor \cite{Shor2} has shown that these conjectures
are globally equivalent to \eqref{Sadd}.   However, the validity of \eqref{Sadd}
for some pair of channels need not imply \eqref{Cadd} for the same pair, or vice versa.  
In  Theorem ~\ref{thm:EBT}, we  
use special features of the channel  to  prove both separatey.   Holevo \cite{HvCC}
proved the more general result that if  \eqref{eofsup} holds for a state associated
with a pair of channels, then it also holds for their conjugates.

Theorem \ref{thm:4} allows one to extend known results
on additivity and multiplicativity  to  their conjugates.  
Conversely, if one can prove these conjectures for some class
of conjugate channels, then one can obtain new results about 
the original class.    The new results obtained thus far  are quite modest.   
This may be partly because the 
conjugate channels   typically   take $M_d \mapsto M_\d2$ with $\d2 > d$,
and channels of this type have not  been studied as extensively.
The next result shows that it would suffice to prove multiplicativity
for a very small and special subset of these maps.
 
\begin{thm}  \label{thm:ext.red}
Suppose that  \eqref{mult} holds for all tensor products $\Psi_1 \ot \Psi_2$
of CPT maps with $\Psi_i: M_{d_i } \mapsto M_{d_i \di}$
whose minimal representation has dimension $\kappa_i \leq \di$.   Then
the multiplicativity conjecture \eqref{mult} holds for {\em all} tensor products
of CPT maps $\Phi_1 \ot \Phi_2$ with $\Phi_i: M_{d_i}  \mapsto M_{\di}$.
\end{thm}
\prf  First, let $\kappa_i$ be the minimum number of Kraus
operators needed to represent $\Phi_i$ and consider the typical case 
$\kappa_i = d_i \di$.  Then $\Phi_i^C: M_{d_i} \rightarrow M_{d_i \di}$, 
but by  \eqref{conj-Kr} requires only $\di$ Kraus operators. 
  Thus, under the hypothesis of the theorem,
\eqref{mult} holds for $\Phi_1^C \ot \Phi_2^C $ .   
But then by Theorem~\ref{thm:4}, \eqref{mult}
also holds for $\Phi_1 \ot \Phi_2$.    When the number of Kraus operators is less than
$d_i \di$, one can simply use a redundant representation with $d_i \di$
operators.   Alternatively,   one could perturb the channel to
$(1 - \epsilon)  \Phi_i +  \epsilon N$, with $N$ the completely noisy map,
and then let $\epsilon \raw 0$. \qed

 Choi showed that a CPT map whose Kraus operators $F_j$ generate a linearly
independent set $F^\dag_j F_k$ in $M_{\d2}$ is an extreme point in the set
of all CPT maps.   Since, $M_{\d2}$ is a vector space of dimension $(\d2)^2$,
this implies that any extreme point can be represented using
at most $\d2$ Kraus operators.     In \cite{RSW} maps which require at most
$\d2$ Kraus operators, but are not true extreme points, are called 
{\em quasi-extreme}
 and all those whose minimal representation  has rank $\kappa \leq \d2$
generalized extreme points.
From a geometric point of view, this extends the extreme points to include some
hyperplanes in regions where the boundary of the convex set of CPT maps
is flat.    For  qubit maps, the quasi-extreme points are
convex combinations of conjugations with two Pauli matrices; these maps 
correspond to edges of the tetrahedron of unital qubit maps.

It was shown in \cite{RSW} that any qubit map which is a generalized extreme 
point has at least two pure output states.    For such maps, \eqref{mult} is trivial.
However, using Theorem~\ref{thm:ext.red} to prove multiplicativity for all qubit
maps, requires proving \eqref{mult} for all maps $\Phi: M_2 \mapsto M_4$
which can be written using two Kraus operators.

For $d > 2$, there are extreme points which do not have pure outputs.  
In particular, when $d = 3$, the Werner-Holevo (WH) counter-example map 
 \cite{WH}Ê is an
extreme CPT map, but all pure states are mapped into projections of rank 2.
(Note that the WH map is not even quasi-extreme when $ d \geq 4$.)
Finding all  generalized extreme points of  CPT maps is a difficult 
problem  which has not yet been solved, even for maps taking $M_3 \mapsto M_3$.
Using \eqref{conj-Kr} it would suffice to find all maps taking 
$M_3 \mapsto M_d$ for $d \leq 9$ whose CJ matrix has rank $\leq 3$.

A non-extreme map $\Phi:M_d \mapsto M_d$ can have minimal dimension $d^2$;
its conjugate $\Phi^C: M_d \mapsto M_{d^2}$ has a   CJ matrix 
which is  $d^3 \times d^3$.   For general maps $\Psi: M_d \mapsto M_{d^2}$ there
are extreme points with minimal dimension $d^2$, but only those with minimal
dimensions $d$ arise from conjugates in this way.   The reduction in
Theorem~\ref{thm:ext.red} to generalized extreme points with CJ matrix
of rank at most $  d$ rather than $d^2$ is quite remarkable.

\section{Conjugates of entanglement breaking maps} \label{sect:EBT}
In this section we review the class of entanglement-breaking maps.
An entanglement-breaking trace-preserving (EBT) map \cite{ShorEBT} is
a CPT map $\Phi$ for   which $(I \ot \Phi)(\rho)$ is separable for all $\rho$.
A number of equivalent criteria are known \cite{HSR}, e.g.,
\be
    \Phi(\rho) = \sum_m R_m \tr E_b \rho
\ee   
where $\{ E_b \}$ is a POVM and each $R_m$ is a density matrix. 
This is the form introduced by Holevo \cite{Hv1}.
Here, we use the
fact that any EBT map  can be written using Kraus operators
$F_k = |x_k \kb w_k |$ with rank one.   Then
\be\label{def:EBT}
     \Phi(\rho) = \sum_k  \proj{x_k} \bra w_k |\, \rho \,| w_k \ket,
\ee
and $\sum_k  \bra x_k | x_k \ket \, \proj{w_k} = I$.  Using the notation
of (\ref{def:F}),
\be
      {\bf F}(\rho) = \sum_{jk} | e_j \kb e_k| \ot |x_j \kb x_k | \,
\bra w_j | \, \rho \, | w_k \ket
\ee
and it follows that
\be  \label{conj.EBT}
    \Phi^C(\rho) =  \sum_{jk} | e_j \kb e_k|   \, \bra x_j | x_k \ket
\, \bra w_j | \, \rho \, | w_k \ket
         =  X* W_{\rho}
\ee
where $*$ denotes the Hadamard product, $X$ is the matrix with elements
$ \bra x_j | x_k \ket $ and $W_{\rho}$ is the matrix with elements
$\bra w_j \, | \rho | \, w_k \ket$, which can be viewed as
a non-standard ``representative'' of $\rho$. Thus in general
the conjugate of an EBT map need not itself be an EBT map.
If we choose $c_{jm}$ such
   that $C C^{\dag} = X$ then the Kraus operators for
(\ref{conj.EBT}) can be written in the form
   \be  \label{pseudodiag}
      R_m = \sum_j c_{j m} | e_j \kb w_j | 
   \ee
where $\{ | e_j \ket \} $ are orthonormal, but $\{ | w_j \ket \} $ need not be.
   Conversely, suppose that the Kraus operators of a map $\Phi$ have the
   form \eqref{pseudodiag} with $\{ e_j \}$ orthonormal.  Then a
straightforward calculation shows that   $\Phi^C$ is an EBT map. 

A special case of  (\ref{def:EBT}) arises
when  the $w_k$ form an orthonormal basis. In this case  $ \Phi$ is 
called a classical-quantum or CQ channel.    Moreover,
  the fact that  $\proj{x_k}$ are rank one implies that $\Phi$ is an  extreme
   point of the set of CPT maps and hence an extreme point of the set of EBT maps
   \cite{HSR}.
  From (\ref{conj.EBT}) it follows that $W_{\rho}$
is the usual matrix representative of $\rho$ in the O.N. basis $w_k$ so that
$\Phi^C(\rho)  = X * \rho$.   Since $X$ is positive semidefinite this
implies that  $\Phi^C$ has simultaneously diagonal Kraus operators,
   and one can easily see from \eqref{pseudodiag} that this is the
   case when the $w_k$ are orthonormal.
This class of  channels was introduced in \cite{LS} where it was called
  ``diagonal''.    We prefer to
call them ``Hadamard channels'' or ``Hadamard diagonal'' 
maps.\footnote{The term ``diagonal channel'' seems to be a natural    
choice for a different class of channels, namely those whose matrix
representative in a particular basis is diagonal.   In Section~\ref{sect:Pdiag}
we consider a class which is seems natural to call  Pauli diagonal channels.}
Thus,    the conjugate of an extreme CQ channel is a Hadamard diagonal  channel.
   King \cite{King1,King5} has shown that both CQ channels and arbitrary
   (not necessarily trace-preserving) Hadamard   diagonal CP maps satisfy
   the multiplicativity \eqref{mult} for all $p$.    Since this holds
   trivially for  both extreme CQ channels and Hadamard diagonal CPT maps,
   we do not obtain a new result.

To get a better understanding of the general case, note
 that an arbitrary  $\d2 \times d$ matrix, or operator 
 $Q:  {\bf C}^d \mapsto  {\bf C}^\kappa$,
  can be written as
\bee
     Q = \sum_{jk} a_{jk} |e_j \kb e^\prime_k | = \sum_j  | e_j \kb w_j | 
\eee
with $| w_j \ket = \sum_k a_{jk} | e^\prime_k \ket $.  Thus, the
 restriction in \eqref{pseudodiag} which distinguishes $ \Phi^C$ from
 an arbitrary channel is that the  vectors  $w_k$  are the same for all
 Kraus operators.    The POVM requirement that 
 $\sum_{k = 1}^{\kappa} \proj{w_k} = I_d$
 in \eqref{def:EBT} implies that $\{ | w_k \} $ are orthonormal when 
 $\kappa = d$; this is precisely the CQ case discussed above.    In the
 general case, we can use Theorem~\ref{thm:4} to obtain the following
 result, which extends King's
results in \cite{King3} to CPT maps with $\kappa > d$.
 \begin{thm}  \label{thm:EBT}
 Let $\Phi_1 : M_d \mapsto M_{\kappa}$ be a CPT map with
 the Hadamard form \eqref{conj.EBT} or, equivalently, a Kraus
 representation of the form \eqref{pseudodiag}.  Then for any CPT map
 $\Phi_2$,   the multiplicativity \eqref{mult} holds for all $p \geq 1$,
 the additivity of minimal ouput entropy \eqref{Sadd}Ê holds,
 and additivity of Holevo capacity \eqref{Cadd} holds.
\end{thm}
\prf  The first part of the theorem
 follows immediately from Theorem~\ref{thm:4}, the fact that any
channel satisfying the hypothesis can be written as the conjugate of an
EBT map, and the fact that EBT maps satisfy \eqref{Sadd} \cite{ShorEBT} 
and \eqref{mult} \cite{King3}.   To prove \eqref{Cadd},  use
\eqref{LSform} to  define $ \gamma_{AC}$ as before \eqref{eof}.
 Since each  $\proj{\psi_j}$ in \eqref{eof} is a pure state,
     $S\big[ \trp_C \proj{\psi_j} \big] = S\big[ \trp_A \proj{\psi_j} \big]$ 
and it follows immediately that
\be
  \eof(\gamma_{AC}) =  S[\Phi(\rho)] -  \chi[\Phi(\rho)]  =  
       S[\Phi^C \! (\rho)] -  \chi[\Phi^C \! (\rho)] 
\ee
If $\gamma_{A_1C_1 A_2 C_2} $ is associated with  a state $\rho_{12}$
using the product representation  \eqref{LSform}  for a pair of
channels and any one of $\Phi_1, \Phi_2, \Phi_1^C, \Phi_2^C$ 
is EBT,   then \eqref{eofsup} holds.
(This result follows immediately from eqn. (25)
in \cite{ShorEBT}, as noted in \cite{DVC};
the same result appears in Lemma 3 of \cite{King2}).    Now let $\rho_{12}$ 
achieve the supremum in
\be
   C_{\hv}(\Phi_1 \ot \Phi_2) = \sup_{\rho_{12}}
      \chi\big[ (\Phi_1 \ot \Phi_2)(\rho_{12}) \big]
\ee
Then, as shown in \cite{MSW}, it follows from \eqref{eofsup} and the
subadditivity of entropy that
\bee
   C_{\hv}(\Phi_1 \ot \Phi_2)   \leq   C_{\hv}(\Phi_1) + C_\hv(\Phi_2)
\eee
Since the reverse inequality is trivial, \eqref{Cadd} holds.  \qed

\section{Conjugates of Pauli diagonal channels}\label{sect:Pdiag}

\subsection{Basic set-up}

In this section we consider a subclass of convex combinations of
unitary conjugations
that can be regarded as the generalization to $d$-dimensions of the
unital qubit channels.

In the case of a  unital qubit channel we can assume, without loss of
generality,
that
$\Phi(\rho) =  \ds{\sum_{k=0}^3} a_k \sigma_k \rho \sigma_k$
where $a_k \geq 0, ~ \sum_k a_k = 1$ and
$\sigma_k$ are the usual Pauli matrices, with the convention that
$\sigma_0 = I$.
One can write a qubit density matrix as
\be
\rho = \half[ w_0 I + {\bf w} \cdot \sigma] = \half \sum_{k=0}^3 w_k \sigma_k.
\ee
where $w_0 = 1 \geq  |\bw|^2 =  \sum_{k=1}^3 w_k^2$.
Then one can choose $F_k = \sqrt{a_k} \, \sigma_k$ and
\be
     \Phi^C(\rho) = \sqrt{A} \pmx
         w_0 & w_1 & w_2 & w_3 \\ w_1 & w_0 & -i w_3 & i w_2 \\
         w_2 & i w_3 & w_0 & - i w_1 \\ w_3 & -i w_2 & i w_1 & w_0 
\emx  \sqrt{A}
          = 4 \sqrt{A}  \, N^C\!(\rho) \,  \sqrt{A}
\ee
where $A$ is the diagonal matrix with elements $ a_j \delta_{jk}$ and
$N^C$
is the conjugate of the completely noisy map for which all $a_k =
\tfrac{1}{4}$.

   To generalize this to  dimension $d > 2$, we first observe that any
orthonormal basis for $M_d$ yields a set of Kraus operators for
the completely noisy channel.   (To see this note that $E_{jk} = |j \kb k|$
is a set of Kraus operators satisfying 
  $\tr E_{ik}^\dag E_{j \ell} = \delta_{ij} \delta_{k \ell}$,
and that any orthonormal basis is unitarily equivalent to $\{ E_{jk} \}$.)
 Let  ${\cal T}$ denote such a basis with the additional requirement
that every element is unitary and the first is the identity, i.e.,
\be   \label{Tdef}
{\cal T} = \{ T_m :  T_0 = I, \tr T_m^\dag T_n = d \delta_{mn},  T_m^\dag T_m =I, ~ m = 0,1 \ldots d^2-1 \}
\ee  
 Then these operators  generate the completely noisy channel via
\be  \label{eq:nois}
    \tfrac{1}{d^2} \sum_{m=0}^{d^2 -1}   T_m \rho T_m^{\dag} =
       (\tr \rho) \,  \tfrac{1}{d} I \equiv   N(\rho) .
\ee
Now consider  channels
\be   \label{Cort}
    \Phi(\rho) = \sum_{m=0}^{d^2 -1}   a_m  T_m \rho T_m^{\dag}
\ee
with $a_m \geq 0, ~ \sum_m a_m = 1$.  The Kraus operators for
this channel are $F_m = \sqrt{a_m} \, T_m$.    One then finds
\be   \label{Pdiag}
   \Phi^{C,T}(\rho)  & = & \sum_{mn} |e_m \kb e_n|  \sqrt{a_m a_n} \,\, \tr
T_m \rho T_n^{\dag}  \nn \\
         & = & d^2 \sqrt{A} \, N^{C,\cT}\!(\rho) \, \sqrt{A} \, = \, d^2 \proj{\alpha} * N^{C,\cT}\!(\rho)
  \ee
where $A$ is the diagonal matrix with elements $a_m \delta_{mn} $,
  $|\alpha \ket$ is the vector with elements $\sqrt{a_m}$, and
we use the superscript $\cT$ to emphasize that  $N^{C,\cT}$, the conjugate
of the completely noisy channel, is constructed using a specific  choice
for the set of Kraus operators.    Thus, the conjugate of a channel of the form \eqref{Cort}
can be written as the composition   $ \Phi^{C,T} = \Psi \circ N^{C,\cT}$
with $\Psi$ a Hadamard diagonal channel with a single Kraus operator, $\sqrt{A}$.

Now define ${\cal N}^\cT = \{ N^{C,\cT}\!(\rho) : \rho = \proj{\psi} \}$ to be the image
of the conjugate of the completely noisy channel acting on pure states.   
Then Corollary~\ref{thm:2} allows us to rewrite the maximal $p$-norm as a variation
over elements of ${\cal N}^\cT$.
\begin{thm}   \label{thm:noise}
 Let $\Phi$ be a channel of the form \eqref{Cort} in the basis ${\cT}$.  Then
  \be  \label{unit4}
     \nu_p(\Phi ) & = & 
     d^2   \sup_{\gamma \in {\cal N^\cT}}  \norm{\sqrt{A} \, \gamma \,
\sqrt{A} }_p    =  d^3 \,  \sup_{\gamma \in {\cal N^\cT} }   \norm{
\gamma   \, A \,  \gamma  }_p                 \ee
  where  
  $A$ is the diagonal matrix with elements $a_m \delta_{mn} $.
  \end{thm}     
\prf For all $\gamma \in {\cal N^\cT}$,  it follows from Theorem~\ref{thm:1} that 
  the non-zero eigenvalues of $\gamma$ are $\frac{1}{d}$ which
  implies that $d \gamma$ is a rank $d$ projection.
 Therefore,
 \bee  \norm{\sqrt{A} \, \gamma \, \sqrt{A} }_p  =
               \norm{\sqrt{\gamma}
\, A \, \sqrt{\gamma} }_p   =
             d    \norm{ \gamma   \, A \,  \gamma  }_p
    \eee
Then \eqref{unit4} follows from Corollary~\ref{thm:2}.    \qed

Despite the apparent simplicity of \eqref{unit4} and the expressions
for $ \Phi^{C,T}$ above, it is not easy to
 exploit Theorem~\ref{thm:noise}.    In order to do so, we need to choose a specific
 basis and obtain more information about
the set ${\cal N^\cT}$.
 
 \subsection{Generalized Pauli bases}   \label{sect:gpb}
 
 We will be particularly interested in bases $\cT$ which satisfy \eqref{Tdef}
 and have  the additional property
that
\be  \label{trip}
       T_m^\dag T_n = e^{i \phi_{kmn}} T_k 
\ee
where $k$ depends on $m,n$.
In this case,  $\tr T_m \rho T_n^{\dag} = e^{- i \phi_{kmn}} \tr T_k \rho$ so that each
row of $N^{C,T}\!(\rho) $ is determined by permuting the elements of the
first row  after multiplication by suitable phase factors.     When $\cT$ has
the property that $T_m  \in \cT \imp T_m^\dag = T_{m^{\prime} }\in \cT$ 
then one can interpret
\eqref{trip} as defining a group operation on $\cT$.

One particular realization of $T_m$ satisfying \eqref{trip} is given by   
 the generalized
Pauli matrices $X^j Z^k,~ j,k = 0 \ldots d \mm 1$ with $T_0 = I$,
and, e.g., $T_m =  X^j Z^k$ for $m = (d-1)j + k$.
Given a fixed
orthonormal basis  $\{ |e_i \ket \}$for ${\bf C}^d$,  the matrices
$X$ and $Z$ can
be defined by
\be\label{def:XZ}
    X |e_k \ket = |e_{k+1} \ket    \qquad \hbox{and}  \qquad   
      Z |e_k \ket = e^{2 \pi i  (k/d)} |e_k \ket 
\ee
with addition mod $d$ in the subscript.   It will then be convenient to
identify $w_{jk} = v_m$ for $m = (d-1)j + k$.    

     When $d = d_1 d_2$, we will also want to 
consider $T_m$ which are tensor products of the generalzied
Pauli matrices, particularly when studying additivity and multiplicativity.
Ritter \cite{Ritt} has considered $T_m$ given by the so-called Gell-mann
matrices which arise in the representation theory of $SU(n)$.

The generalized Pauli matrices satisfy the commutation relation
\be
    ZX = e^{i 2 \pi/d} XZ.
\ee
It then follows that the matrix representing a channel $\Phi$ of the form
\eqref{Cort} in this basis,
is diagonal.  In fact
\be
     \tr (X^i Z^k)^\dag \Phi(X^j Z^{\ell} )= \delta_{ij} \delta_{k \ell} \lambda_{jk}
\ee
with  
\be  \label{Plamb}
    \lambda_{jk} = e^{jk \frac{ 2 \pi i}{d} } \sum_{mn}  e^{ (mk-jn) \frac{i2 \pi i}{d}  } a_{mn} 
        =      \ovb{\lambda}_{d-j,d-k}.
    \ee
Moreover,  
\be\label{def:lambda}
  \Phi: \td \big[I + \sum_{jk} w_{jk} X^j Z^k\big] \mapsto   
     \td \big[I + \sum_{jk}  \lambda_{jk} w_{jk} X^j Z^k\big] .
 \ee
 We will call channels of the form \eqref{Cort} in the generalized
 Pauli basis {\em Pauli diagonal}  channels.    They are a natural
 generalization of the unital qubit channels.     Pauli diagonal
 channels are Weyl covariant which implies \cite{Cort,Hv2}
 \be   \label{weyl}
     C_\hv(\Phi) = \log d - S_{\min}(\Phi).
 \ee

 Any channel  $\Psi$ can be represented in a basis $\cT$ by the matrix
 $X$ with elements  $x_{mn} = \tr T_m^\dag \Psi(T_n)$.    
When $\Psi$ is trace-preserving, $x_{0n} = \delta_{0n}$ and
 if $\Psi$ is unital $x_{m0} = \delta_{m0}$.     In the case of qubits, any
 unital channel can be  diagonalized in the usual Pauli basis by using 
 the singular  value decomposition and the correspondence between
 rotations in ${\bf R^3}$ and unitaries in $M_2$.  (See \cite{KR1} for details.)   
 One could, in principle, use the singular  value decomposition to
 diagonalize $X$.   However, the corresponding change of bases
 will not normally preserve the properties \eqref{Tdef} and \eqref{trip}.
    Thus, even a convex combination of conjugation
 with arbitrary unitary conjugations can not necessarily be written in
 diagonal form using the generalized Pauli basis.     In fact, when $d = d_1 d_2$,
 a channel which is diagonal in a tensor product of Pauli bases  
 need not be diagonal  in the generalized Pauli basis for $d$, and
 vice versa.   (This is easy to check for $d = 4$, $d_1 = d_2 = 2$.)
 
In at least one non-trivial case it is possible to explicitly compute the maximal
$p$-norm of this class of channels, that is when $p=2$ and $d=3$.
 
 \begin{prop}   \label{prop:mike}
The maximal 2-norm of a Pauli diagonal channel satisfies the bound
\be   \label{nud3p2}
    \nu_2(\Phi) \leq  d^{-1/2} \Big( 1 + (d - 1) \sup_{(j,k) \neq (0,0)} |\lambda_{jk}|^2 \Big)^{1/2}
\ee
where $\lambda_{jk}$ is given by \eqref{Plamb}.   When $d = 3$, the bound is attained
for a state of the form $\tfrac{1}{3} \big[ I + X^{j_*} Z^{k_*} +  (X^{j_*} Z^{k_*} )^2 \big]$
where $j_*, k_*$ denote the pair of integers for which the supremum is attained
in \eqref{nud3p2}.
\end{prop}
\prf   Using the notation of (\ref{def:lambda}),
\be   \label{p2mike}
   \norm{\Phi(\rho)}_2^2 & = & \tr [\Phi(\rho)]^\dag  \Phi(\rho)  \nn \\ \nn 
         & = & \tfrac{1}{d^2} \sum_{ikj\ell} \ovb{\lambda}_{ik}  \ovb{w}_{ik} \lambda_{j\ell} w_{j\ell}  \,
           \tr  Z^{-k} X^{-i} X^j Z^{\ell}  \\ \nn 
                  & = &  \td \sum_{jk}     | \lambda_{jk}|^2 |w_{jk}|^2  \\ \nn 
        &   \leq &  \td \Big[ 1 +     \sup_{(j,k) \neq (0,0)}   | \lambda_{jk}|^2 
           \sum_{(j,k) \neq (0,0)} |w_{jk}|^2 \Big] \\
      & = & \td \Big[ 1 +    (d-1) \sup_{(j,k) \neq (0,0)}   | \lambda_{jk}|^2 \Big]
\ee
with $\lambda_{jk}$  given by \eqref{Plamb}.
One can then verify that the bound is attained with the indicated state.  \qed

Fukuda and Holevo  \cite{FH} independly proved the inequality \eqref{nud3p2}.
Moreover,  when equality holds for some channel $\Phi_1$, 
 then the multiplicativity conjecture \eqref{mult} holds for $p = 2$
 with $\Phi_2$ any other CPT map.   In Example~\ref{sect:axes}, we show that that equality
 holds for a special class of Pauli diagonal channels.

\subsection{Representations of density matrices} \label{sect:DM}

Since $\cT$ is an orthonormal basis for $M_d$, any density matrix 
can be written as
\be   \label{eq:bloch}
     \rho = \frac{1}{d}\Big[I +  \sum_{m = 1}^{d^2-1}  v_m T_m \Big] ,
\ee
with $v_m = \tr T_m^\dag \rho$.    This implies $|v_m| \leq \norm{T_m} \tr \rho = 1$.
However, finding conditions on
$v_m$ which ensure that an expression of the form \eqref{eq:bloch}
 is positive semi-definite is far from trivial. 
When $\rho$ is a  pure state,
 $1 = \ds{\tr \rho^2 = \tfrac{1}{d^2} \big[1 + \sum_{m = 1}^{d^2-1}  |v_m|^2 \big]}$,
 so that
\be  \label{rowsum}
 \sum_{m = 1}^{d^2-1} |v_m|^2 = d - 1.
 \ee
Combining this with $|v_m| \leq 1$ implies that every pure state has at least
$d$ non-zero coefficients (including $v_0 = 1$).
For mixed states, one can have fewer non-zero coefficients.   For example,
when $d = 4$,  $\rho = \tfrac{1}{4} [I + Z^2]$.

 For  $\rho = \proj{\psi} $   a pure state written in the form \eqref{eq:bloch},
  define $  {\cal S} $ as the subgroup generated by  $\{ T_m : v_m \neq 0 \}$.    
    It follows from the fact that at least $d$ coefficients are non-zero
 that any subgroups generated by a pure state in this way have
 $|  {\cal S}| \geq d $.

  In the generalized Pauli basis, two examples of ${\cal S}$ are
  $ \{ I, X, X^2, \ldots X^{d-1} \} $ and $\{ I, Z, Z^2, \ldots Z^{d-1} \} $.
  In fact, any choice of  $W = X^j Z^k $ 
 with $j$ or $k$ relatively prime  to $d$ 
  generates a cyclic subgroup 
    \be  \label{cycsub}
    {\cal S} = \{ I, X^j Z^k, (X^j Z^k)^2, \ldots, (X^j Z^k)^{d-1} \},
  \ee
  and the projections onto orthogonal eigenvectors  of $W = X^j Z^k $ can be written as
  \be   \label{axis}
      \proj{\psi_n} = \td \big[ I + \sum_{j=0}^{d-1} \omega^{nj} W^j \big]   \qquad n = 1 ,2 \ldots d
  \ee
 with $\omega = e^{2 \pi i /d}$.    We will call such states {\em axis states}.  When $d$  is prime,
 there are
 $d+1$ distinct subgroups of the form \eqref{cycsub}, whose eigenvectors
  generate $d+1$ orthogonal bases for ${\bf C}^d$.   
 These are the $d+1$ mutually unbiased bases.
      
      \begin{exam}
{\em    When $d = 4$,
  $  {\cal S} = \{ I, X^2, Z^2, X^2 Z^2 \}$.  In this case, the elements of $  {\cal S}$
  do not commute (although the group is formally abelian) and do not have
  simultaneous eigenvectors.   However, $|\psi \ket = (1,0,1,0)$ satisfies
  \be
      \proj{\psi} =  \tfrac{1}{4} \big[ I + Z^2 + X^2 + X^2 Z^2 \big]
  \ee 
 and  $N^{C,\cT}\!( \proj{\psi} )$ is decomposable.  }  \end{exam}
  
   \begin{exam}
 {\em   When $d = 4$,
  $  {\cal S} = \{ I,  Z^2, X, X Z^2, X^2, X^2 Z^2,  X^3, X^3 Z^2 \}$.  is another
  subgroup, which has order $2d$.   For  $|\psi \ket = (a,b,a,b)$
   \be
      \proj{\psi} =  \half \big[(a^2 + b^2)( I + X^2) + (a^2 - b^2) Z^2 ( I + X^2)   X^2 Z^2
       +   2ab X ( I + X^2)  \big] .
  \ee 
 Note that this pure state does not require the full subgroup, i.e.,
 the coefficients of $X  Z^2$ and $X^3 Z^2 $ are zero.    
(This can not happen for subgroups of order $d$.)
  The terms   $X  Z^2$ and $X^3 Z^2 $ do arise
 in the product $\rho^2$, but since $ab(XZ^2 + Z^2 X) = ab(X Z^2 - X Z^2)  = 0$
 the coefficients are zero.
   } \end{exam}

 \begin{exam} \label{sect:axes}
{\em Let $W_L (\nu = 1,2 \ldots d \pp1)$ denote a set of fixed generators for $\kappa$ cyclic  groups
of the form \eqref{cycsub}, chosen so that the groups are mutually
disjoint except for the identity.    Let
\be   \label{con.axis}
    \Omega = s \id + \sum_{L = 1}^{\kappa}    t_L \Psi_{L}^{\rm QC}+ u {\cal N}   
\ee
where  $\Psi_L^{\rm QC} $ is the channel that
maps a state $\rho$ onto its diagonal when it is written in the axis basis 
\eqref{eq:bloch} for $W_L$.   The condition
$s + \sum_L t_L + u = 1$  implies that $\Omega$ is trace-preserving, and the
conditions
\be  \label{axisco}
   a_0 = s +  \td \sum_L t_L + \tfrac{1}{d^2} u \geq 0, \qquad   
          a_L \equiv \td t_L  + \tfrac{1}{d^2} u \geq 0
\ee
are necessary and sufficient for  $\Omega$ to be CP.   With
the correspondence  $T_m \sim X^j Z^k \sim  W_L^{n}$,  the
 coefficients in \eqref{Cort} depend only on $L$ and are given
 by \eqref{axisco}.  Moreover, the parameters in \eqref{Plamb} also depend
 only on $L$ and satisfy
 $ \lambda_{jk} \sim  \lambda_L = s + t_L$.    Let 
 $\lambda \equiv \max _L | \lambda_L|   \equiv \max _L | s + t_L|$.
 One can verify that $[\nu_2( \Omega)]^2 = \td [1 + (d-1)\lambda^2]$ is attained
 with the axis states \eqref{axis} for  an $L$ which attains $\lambda$.      
 Thus, by Theorem~2  
 in  \cite{FH},   multiplicativity  
  \eqref{mult} holds for   $\Omega \ot \Phi$ when  $p = 2$ and   $\Phi$ is any CPT map.     
  The case of only one non-zero $t_L$ was also considered in   \cite{FH}.     
  Extensions to mutually unbiased bases when $d$ is a prime power are
  considered in \cite{Rusk}.}
  \end{exam}

\subsection{Image of the completely noisy conjugate}

The coefficients $v_m$ form the first row of the matrix $N^{C,T}\!(\rho) $ 
  so that $\rho \neq \gamma$ implies  $N^{C,T}\!(\rho) \neq N^{C,T}\!(\gamma) $.
This uniqueness allows one to consider $N^{C,T}\!(\rho)$ as a representation
of the set of density matrices, and \eqref{eq:bloch} might be regarded as a
generalization of the Bloch sphere representation.     Indeed, if the 
(non-unitary standard basis)
is ordered so that $T_{j +(k\mm1)d} = E_{jk} = |j \kb k|$, then
$N^{C,T}\!(\rho) =   \td I_d \ot \rho $.   
Combining this observation with Lemma~\ref{lemma1} gives 
\begin{thm}   \label{thm:bob}
For any basis $\cT$ satisfying \eqref{Tdef}, there is a unitary matrix $U_{\cT}$
such that $N^{C,T}\!(\rho) =   U_\cT  \,\td  I_d \ot \rho  \, U_\cT^\dag$.
\end{thm}

This result has an interesting interpretation with potential applications.  
It says, in the terminology of the introduction, that one can actually use
noise to transmit information for Alice to Bob.   In fact, when the noise
has completely destroyed Alice's information  (i.e., her density matrix
is $\td I$), Bob has a faithful copy.   This may be counter-intuitive
because his density matrix also has entropy at least $\log d$.  However,
Bob's system has dimension $d^2$ and can be regarded as itself a
composite of two d-dimensional subsystems $B_1$ and $B_2$.  
Theorem~\ref{thm:bob} implies that Bob can make a unitary transformation
on his system so that all the noise is in one room and a faithful copy of
Alice's original quantum state in the other.     Note that this result applies
to mixed, as well as pure, inputs.

Combining Theorem~\ref{thm:bob} with \eqref{Pdiag}
 gives the following
\begin{cor}  \label{cor:bob}
The conjugate of a Pauli diagonal channel can be written as
\be
\Phi^{C,P}(\rho) = \sqrt{A} \, U_\cP  \, \td I_d \ot \rho  \, U_\cP^\dag \sqrt{A} = F  \, I_d \ot \rho  \, F^\dag 
\ee
where,  $U_\cP $ is the  unitary matrix which transforms the standard basis
$\{ E_{jk} \}$ to the generalized Pauli basis, 
 $A$ is a positive diagonal operator with $\tr A = 1$,
and $F = d^{-1/2} \sqrt{A} \, U_\cP  $.
\end{cor}
This is  essentially the Stinespring representation for $\Phi^{C,P} \!Ê(\rho)$.
Since  $\Phi^{C,P} \! (\rho) $ is trace-preserving, 
$\trp_1 \, F^\dag F = \trp_1 \, \td U_\cP^\dag A U_\cP = I_2$.   A similar
result holds for other channels which are diagonal with respect to a
set of unitary Kraus operators.

 \begin{thm}   \label{thm:Nc}
For any pure state $\proj{\psi}$, the state $N^C\!(\proj{\psi})$ satisfies
the following conditions:
\begin{itemize}
\item[a)]  $d \, N^{C,\cT} \!(\proj{\psi})$ is a projection of rank $d$, and
\item[b)]  all diagonal elements of $  N^{C,\cT}!\!(\proj{\psi})$
equal $\tfrac{1}{d^2}$.
\end{itemize}
\noindent In addition, if $T_m$ 
satisfies (\ref{trip}) then
\begin{itemize}
\item[c)] all  elements of $  N^{C,\cT}\!(\proj{\psi})$
are $ \leq \tfrac{1}{d^2}$, and
\item[d)]  $ d^3 N^{C,\cT}\!(\proj{\psi}) * \overline{N^{C,\cT}!\!(\proj{\psi})}$
is a double stochastic matrix.
\end{itemize}
\end{thm}

Theorem \ref{thm:Nc} provides a set of necessary conditions for a
matrix in $M_{d^2}$ to be $N^{C,\cT}\!(\proj{\psi})$ for some pure state.
However, there are matrices in $M_{d^2}$ which satisfy (a), (b), (c), (d) above,
but can {\em not} be realized as the image  $N^{C,\cT}\!(\proj{\psi})$ of
any pure state density matrix.

 A particularly interesting subset of ${\cal N}$ consists of those
for which exactly $d$ of the $w_m$ have $|w_m| = 1$ and the
rest are zero.   When the operators (\ref{def:XZ}) are used,  $N^C\!(\rho)$ is
  permutationally equivalent to a block diagonal matrix with
$d \times d$ blocks on the diagonal, each of which is rank one
and has all elements with magnitude $1$.     We will call such
$N^C\!(\rho)$ {\em $d$-decomposable}.   (In general, a 
decomposable matrix is one which is permutationally equivalent 
 to a block diagonal matrix).    Theorem~\ref{thm:Nc} implies that
 all decomposable matrices in ${\cal N}^{\cT}$ have blocks of the
 same size.
 
Let $  {\cal S} $   be the subgroup  of $\cT$ associated with a pure
state as in Section~\ref{sect:DM}, or, equivalently, generated by the
non-zero elements of the first row of $N^{C,\cT}\!(\proj{\psi})$.
The cosets   $T_k{\cal S}$ define a partition of the integers $\{ 0, 1, \ldots d^2-1 \}$.
   Moreover, if $T_m$ satisfy \eqref{trip}, and $|  {\cal S} | < d$, then
  $N^{C,\cT}\!(\rho)$ is decomposable and the decomposition into blocks
  corresponds to the partition determined by the cosets of ${\cal S}$.  When the
  order of  ${\cal S}$ is $d$, it
  follows from Theorem~\ref{thm:Nc} that each block is a rank 1 projection
  with diagonal elements $\td$; this implies that all of the non-zero coefficients
  satisfy $|v_m| = 1$.

   
   \begin{thm}   \label{thm:prod-ME}
  Let $d$ be prime and $\cP\ot \cP$   the basis for $M_{d^2}$ consisting of
  tensor  products of generalized Pauli matrices.   If $N^{C, \cP \ot \cP}(\proj{\psi})$
  is $d^2$-decomposable, then $|\psi \ket$ is either a product state or a
  maximally entangled state.  
   \end{thm}
   \prf  First observe that  for an arbitrary $|\psi \ket \in {\bf C}^{d^2} \simeq  {\bf C}^d \ot {\bf C}^d$
   \be  \label{eq:doub}
      \proj{\psi} = \tfrac{1}{d^2} \sum_{mn} c_{mn} T_m \ot T_n.
   \ee
When $N^{C, \cP \ot \cP}(\proj{\psi})$ is $d^2$-decomposable, at most
   $d^2$ of the $d^4$ coefficients $c_{mn}$ are non-zero, and the
   corresponding $T_m \ot T_n$ generate a subgroup of order at most $d^2$.
   This implies that \eqref{eq:doub} must reduce to one of the following
   two forms.
   \begin{align}   \label{ddec1}
  \gamma_{12} =    \proj{\psi_{12}} & = \tfrac{1}{d^2}  \Big( I \ot I + 
      \sum_{m=1}^{d^2-1}  e^{i \theta_m} T_m \ot T_{\pi(m)} \Big) \\
 \intertext{  where $\pi$ is a permutation of  $\{ 1,2 \ldots d^2-1 \}$, or,}
    \label{ddec2}
  \gamma_{12} =   \proj{\psi_{12}} & = \tfrac{1}{d^2}  \sum_{m =0}^{d^2-1} \sum_{n =0}^{d^2-1}
     e^{i \theta_{mn}} V^m \ot W^n
 \end{align}
 where  $V = X^i Z^k$ for some fixed $i,k$ and $W = X^j Z^\ell$ for some fixed $j,\ell$.
 
 In the first case \eqref{ddec1} we have used the fact that the trace-preserving 
 property requires the  term  $I \ot I$ and the requirement of a group of order $d^2$
 implies that once one goes beyond a cyclic subgroup each $T_m$  can only
 occur once.   In this case, it is immediate that $\gamma_1 = \gamma_2 = \td I$
 which implies that $\psi_{12}$ is maximally entangled.
 
 In the second case \eqref{ddec2}, the subgroup is a direct product of cyclic subgroups.
The requirement that $| \psi_{12} \ket$ is pure is equivalent to  
\be  \label{ddec3}
e^{i \theta_{mn} }  =    \tfrac{1}{d^2} \sum_s \sum_t    e^{i \theta_{st} } e^{i \theta_{m-s,n-t} }
 \ee
  with subscript addition mod $d$.    It then follows fromÊthe triangle inequality that
  \be
      1 &  \leq &  \tfrac{1}{d^2} \sum_{st} 1  = 1,
  \ee
 which implies     $e^{i \theta_{m-s,n-tn}} = e^{i \theta_{mn}}  e^{-i \theta_{ st}}$

Now, since  $\tr V^m = d \delta_{m0}$,
  $\gamma_1 = \tr   \gamma_{12} = \td \sum_m e^{i \theta_{m0} }V^m$,
  and the condition that    $\gamma_1$ is pure is 
  $e^{i \theta_{m0} } = \td \sum_s  e^{i \theta_{s0} }  e^{i \theta_{m-s,0} } $.
  But this holds,  since we have already shown that  
   $e^{i \theta_{m-s,0} } = e^{i \theta_{m 0} } e^{-i \theta_{s0} }$.
   Therefore,   $\rho_1$ is a pure state $\proj{\psi_1}$; similarly
   $\rho_2 = \proj{\psi_2}$.   Since $\rho_{12}$ is pure, this implies that 
$| \psi_{12} \ket = |\psi_1 \ket \ot  |\psi_2$  is a product.   \qed
     
We conclude this section with an explicit expression for $N^{C,\cT}( \proj{\psi})$ in
the generalized Pauli basis.
\begin{thm}   \label{thm:NcP}
In the generalized Pauli basis,
\be   \label{eq:NcP}
     N^{C,\cP}( \proj{\psi}) = \sum_{\ell} X^\ell  R \proj{\psi} R X^{-\ell}  \ot Z^\ell
        \proj{\iota} Z^{-\ell}  
\ee
where $|\iota \ket $ is the vector whose elements are all $1$ and 
$\ds{ R \sum_k v_k | k \ket  =    \sum_k v_{d-k} | k \ket }$, i.e., $R$  reverses the order
of the elements of a vector.
\end{thm}
As an immediate corollary, we find that
\be
     {\cal N}^{C,\cP} =   \Big\{ \sum_{\ell}  X^\ell    \proj{\psi}   X^{-\ell} \ot Z^\ell \proj{ \iota} Z^{-\ell}  
               :  \psi \in {\bf C}^d \Big\}
\ee
\prf   By  a straightforward calculation one finds
\bee 
  N^{C,\cP}( \proj{\psi}) & = & \sum_{jk} \sum_{mn}   |j \ot m \kb k \ot n|  \tr 
      X^j Z^m \proj{\psi} (X^k Z^n)^\dag \\
      & = & \sum_{jk} \sum_{mn}  \sum_\ell  \ovb{\psi}_{\ell  } \psi_{\ell  } 
          \omega^{(m-n) \ell}   \bra   \ell +j,     \ell +k \ket
          |j \ot m \kb k \ot n|   \\
      & = &  \sum_\ell  \Big(   \sum_{jk}      \ovb{\psi}_{\ell - j}   \psi_{\ell - k}  | j \kb k| \Big)
    \bigotimes    \Big(  \sum_{mn} |m \kb n | \Big)     \omega^{(m-n) \ell} 
         \eee
         where $\omega = e^{2 \pi i/d}$.  \qquad \qed
\newline
Note that the last line   says that  each block of $N^{C,\cP}( \proj{\psi}) $
is cyclic, and the expression is
   consistent with the fact that the first row determines
the rest.

\subsection{Upper bound on $\nu_p(\Phi)$}

\begin{thm}  \label{thm:mike}
   For a channel of the form \eqref{Cort}, let  $b_j$ be
a rearrangement of $a_j$
   in non-increasing order, and define $\beta_j = \sum_{i =
0}^{d -1}  b_{i+jd} , 0 \le j \le d-1$. Then
\be  \label{upbd}
     \nu_p(\Phi) =  \nu_p(\Phi^{C,\cT}) \leq   \Big( \sum_{i = 0}^{d-1} \beta_i^p \Big)^{1/p}
   \ee
Moreover, if equality holds, $\Phi^{C,\cT}\!(\rho)$ is decomposable
for a $\rho$ that maximizes the $p$-norm.
\end{thm}
\prf
By (\ref{unit4}), it suffices to bound
$\vert\vert \gamma A \gamma \vert\vert_p$ for $\gamma
\in  {\cal{N^\cT}}$.      Every eigenvector of $\gamma A \gamma$
corresponding to a non-zero eigenvalue is in the range of 
$\gamma$.     Therefore, we can choose an orthonormal 
basis for the range of  $\gamma$ consisting of normalized
eigenvectors $|f_i\ket$ of $\gamma A \gamma$ arranged in order of
non-increasing eigenvalues $\lambda_i$.  (It may
be necessary to include some eigenvectors with eigenvalue
zero.)  By    Theorem~\ref{thm:Nc},  
$d \gamma$ is projection of rank $d$; therefore, we can write
\be
\gamma = \frac{1}{d} \sum_{i = 0}^{d-1} \vert f_i \kb f_i \vert.
\ee
Since $A$ is diagonal with elements $a_r \delta_{rs}$ in the
standard basis $\{ |e_r \ket$, we find 
\be
   \lambda_i = \bra f_i \vert \gamma A \gamma \vert
f_i \ket
   =  \frac{1}{d^2} \bra f_i \vert A \vert f_i \ket
= \frac{1}{d^2}\sum_{r = 0}^{d^2-1}  a_r  \vert
\bra f_i \vert  e_r \ket \vert^2 \label{lambdadef}
\ee
for $i = 1 \ldots d$.    The inequality \eqref{upbd} will follow 
from standard results \cite{HJ1,MO} if  we can show that the
 eigenvalues of $d^3\gamma A \gamma$ are majorized by $\{ \beta_i \}$.

By part (b) of Theorem~\ref{thm:Nc} , 
\be \label{diagonalbound}
\sum_{i} \vert \bra f_i \vert e_r \ket
\vert^2 =  d\bra e_r \vert \gamma \vert e_r \ket =
\frac{1}{d} .
\ee
Since $\vert f_i \ket$ is a unit vector, we also have
$\sum_r \vert \bra f_i \vert  e_r \ket \vert^2 = 1$.  
Therefore
\be   \label{eqsum}
\sum_{i = 0}^{k-1} \sum_{r = 0}^{d^2-1}
    \vert \bra f_i \vert  e_r \ket \vert^2  \, = \, k \, = \,
\sum_{i = 0}^{d-1} \sum_{r = 0}^{kd-1}
    \vert \bra f_i \vert  e_r \ket \vert^2  .
    \ee
Removing the common terms 
  $ \sum_{i = 0}^{k-1} \sum_{r = 0}^{kd-1} \vert \bra f_i \vert  e_r \ket \vert^2$
 in (\ref{eqsum}) gives the identity
\be\label{eqsum1}
\sum_{i = 0}^{k-1} \sum_{r = kd}^{d^2-1}
    \vert \bra f_i \vert  e_r \ket \vert^2  \, = \, 
\sum_{i = k}^{d-1} \sum_{r = 0}^{kd-1}
    \vert \bra f_i \vert  e_r \ket \vert^2  .
    \ee
Since  $b_r $ is a rearrangement of $a_r$, we can assume
without loss of generality that the basis $|e_r \ket $ has been chosen
to correspond to the ordering of $b_r$.   Then   
$s < kd < t$ implies $b_s \geq b_{kd} \geq b_t$ and it follows
that for each $k = 1,2, \ldots d \mm 1$.
\be
d^3 \sum_{i = 0}^{k-1} \lambda_i & = &
  d \sum_{i =
0}^{k-1} \sum_{r = 0}^{d^2-1}  b_r  \vert \bra f_i
\vert  e_r \ket \vert^2 \label{eigenv} \\  \nn 
& = & d \sum_{r = 0}^{kd-1}  b_r  \Big(\sum_{i =
0}^{k-1}  \vert \bra f_i \vert  e_r \ket \vert^2 \Big) +
d \sum_{r = kd}^{d^2 -1}  b_r  \Big(\sum_{i =
0}^{k-1}  \vert \bra f_i \vert  e_r \ket \vert^2 \Big) \\ \nn 
& \leq &
d \sum_{r = 0}^{kd-1}  b_r  \Big(\sum_{i =
0}^{k-1}  \vert \bra f_i \vert  e_r \ket \vert^2 \Big) + d \, b_{kd} \,
\sum_{i = k}^{d-1} \sum_{r = 0}^{kd-1}
    \vert \bra f_i \vert  e_r \ket \vert^2 \\ \nn 
& \leq &
d \, \sum_{r = 0}^{kd-1}  b_r \Big(\sum_{i =
0}^{d-1}  \vert \bra f_i \vert  e_r \ket \vert^2 \Big) \\
& \le& \sum_{r = 0}^{kd-1} b_r \label{majorize}
   =  \sum_{j = 0}^{k-1} \beta_j
\ee
  where we used \eqref{eqsum1} for the first inequality.
  For $k = d$, it follows immediately from \eqref{diagonalbound} and
  \eqref{eigenv} that 
  \bee
  d^3 \sum_{i = 0}^d \lambda_i =  \sum_{r=0}^{d^2 - 1} b_r = \sum_{j = 0}^{d} \beta_j
  \eee
  Thus, the eigenvalues of $d^3\gamma A \gamma$ are majorized $\{ \beta_i \}$.
If $\gamma$ is not decomposable, then for some $r$,
\be
0 < \vert \bra f_1 \vert e_r \ket \vert^2 <
\sum_{i=0}^{d-1} \vert \bra f_i \vert e_r \ket \vert^2
= \frac{1}{d^2}
\ee
which implies  a strict inequality in (\ref{majorize}).
  \qquad \qed
  
  Note that the numbers $\beta_j$ define a partition of the integers 
  $\{ 0, 1, \ldots d^2-1 \}$.    Let  $0, m_1, m_2, \ldots m_{d-1}$ be the
  subset which contains $0$, and let   
    \be
    {\cal S }   =   \{ I, T_{m_1} , T_{m_2} , \ldots, T_{m_{d-1} } \} .
  \ee 
  If the set $ {\cal S }$ is not a  subgroup of $\cT$ then there
  is no pure state $\rho$ which will generate a $d$-decomposable 
  $\gamma \in {\cal N}^\cT$ for which the  upper bound \eqref{upbd} is attained.  

  \subsection{Applications to Multiplicativity}
    
To use Theorem~\ref{thm:mike} to prove multiplicativity \eqref{mult}, one
would need to show that the upper bound \eqref{upbd} is attained for
both $\Phi$ and $\Phi \ot \Phi$.  
   Unfortunately, this is almost never true for  $\Phi \ot \Phi$, 
and for the few exceptions multiplicativity is well-known
  and easily proved.   One can, however, prove a new result for the
  $p = \infty$ norm.   Before doing so, we present two insightful examples.

When   a channel has the form \eqref{Cort} in the generalized Pauli basis,
we use slight abuse of notation and write $a_{jk}$ for the weight given to conjugations
with $X^j Z^k$. 
\begin{exam}  \label{ex:PQC} {QC channels:}
{\em If $a_{jk} = a_j$  does not depend  on $k$, then $\beta_j = d a_j$ and it is
easy to see that the upper bound \eqref{upbd} can be attained and the corresponding
channel is multiplicative in the sense  
\be
      \nu_p(\Phi^{\ot m}) = \big[ \nu_p(\Phi) \big]^m \qquad \forall ~ p \geq 1 \quad \hbox{and}
      \quad  \forall  ~\hbox{integers} ~ m .
      \ee
However, this does not lead to a new result because
\be   \label{QCsum}
    \sum_{jk} a_j X^j Z^k \rho (X^j Z^k )^\dag =
      \sum_j a_j X^j  \Big( \sum_k Z^k \rho  Z^{-k} \Big) X^{-j} =  \sum_j a_j X^j \rho_{\rm diag} X^{-j}.
\ee
The map $\rho \mapsto \rho_{\rm diag}  $ is a special type of EBT channel called quantum-classical (QC).    Therefore, \eqref{QCsum} is a an EBT map.
} \end{exam}

\begin{exam}    \label{ex:dep}  
{\em  The depolarizing channel is defined as
 \be   \label{dep}
   \Phi(\rho) = b \rho +  \frac{1-b}{d} \tr \rho \, I ,
   \ee
For this channel,  $\nu_p(\Phi)$ is easily
computed and known to satisfy $\nu_p(\Phi \ot \Phi) = [\nu_p(\Phi)]^2$
When $b > 0$, $\Phi$ can be written in the form  \eqref{Cort} with  $a_0 > a_j$
and $a_j = \frac{1-a_0}{d^2 -1} = (1-b)\tfrac{1}{d^2}$ for $j \geq 1$.   
  The upper bound can be attained with a 
decomposable state.      
However, the tensor product $\Phi \ot \Phi$  does {\em not}  
attain the upper bound in the basis given by tensor products of
 generalized Pauli matrices.     To see why, observe that in this
 product basis
 $\beta_0 = a_0^2 + (d^2 \mm 1) a_0 a$.  But it is known \cite{FH,King3}
that this channel is multiplicative for all $p$, which implies that  its largest
eigenvalue is $\big[b  +  \frac{1-b}{d} \big]^2 =  [a_0 + (d \mm 1)a]^2$.
Since   
$$ [a_0 + (d \mm 1)a]^2  = 
     a_0^2  +  2(d \mm 1)a_0 a + (d \mm 1)^2 a^2  < a_0^2 + (d^2 \mm 1) a_0 a = \beta_0 $$
the upper bound is not attained.   Although the product density matrix which attains
$[\nu_p(\Phi)]^2$ can be chosen to be decomposable, its blocks do not correspond
to a partition which attains the upper bound.   

When $- \td \leq b < 0$, one
has $a_0 < a_j$ for $j \geq 1$, but  a similar analysis shows that the
upper bound  is not attained for $\Phi \ot \Phi$. } \end{exam}

The problems which arise in the depolarizing channel are generic.  This is
most easily seen by examining the qubit case in detail, which is done in
Appendix B.    The most one can hope to obtain is the following result
for the infinity norm.

\begin{thm}\label{thm:p=infty}
Let  $\Phi$ be a Pauli diagonal channel and $b_{jk}$ a rearrangement
of $a_{jk}$ as in Theorem~\ref{thm:mike} so that $\beta_j = \sum_k  b_{jk}$.
Let $j_{*}$ denote the index for which $a_{00} = b_{j_{*}k}$ for some $k$,
and
\be  
   {\cal S} = \{ X^m Z^n :  a_{mn} = b_{j_{*}k}, k = 0, 1 \ldots d \mm 1 \}.
\ee
Then the upper bound \eqref{upbd} is attained if and only if ${\cal S}$ is a subgroup
of $\cT$ and its partition into cosets corresponds to
the partition defined by the $\beta_j$, i.e., each coset has the form
\bee
    T_\ell {\cal S} =  \{ X^m Z^n :  a_{mn} = b_{jk}, k = 0, 1 \ldots d \mm 1 \} \quad \text{for some} ~ j.
\eee
If, moreover, $b_{0 , d \mm -1}^2 > b_{00} \, b_{10}$ (under the assumption 
$b_{jk} \geq b_{j, k + 1}$),  then
\be
\nu_\infty(\Phi \ot \Phi) =  \big[\nu_\infty(\Phi)\big]^2.
\ee
More generally, if  $b_{0 , d \mm -1}^r > b_{00}^{r-1} b_{10}$ , then
$\nu_\infty(\Phi ^{\ot r}) =  \big[\nu_\infty(\Phi)\big]^r$.
\end{thm}
\prf   The first part is essentially a matter of notation and our earlier
discussion about subgroups and partitions.   For the second part, it
suffices to observe that the inequality implies that the largest
$\beta$ for $\Phi \ot \Phi$ is $\beta_0^2$.  \qed

\section{Giovannetti-Lloyd linearization operators}\label{GL}

In \cite{GL} a  linearization of $p$-norm functions
was introduced  and subsequently used \cite{GL2,GLR}
used to prove multiplicativity
for integer $p$ and certain special types of channels.
For  any integer $p$, it is possible to find a linear operator
$X(\Phi, p)$ in ${\cal H}^{\otimes p}$ such that
\begin{align}
   \tr(\Phi(\rho))^p=\tr(\rho^{\otimes p}X(\Phi,p))
\label{linearize}
\end{align}
holds for any $\rho$.
   $X(\Phi, p)$ is not uniquely defined.  Initially \cite{GL,GL2},
the realization $ \Theta(\Phi,p)$, defined in terms of the
Kraus operators $A_k$ of $\Phi$ as
\be   \label{theta}
   \Theta(\Phi,p)
= \sum_{k_1,\cdots,k_p}
A^\dagger_{k_1}A_{k_2}\otimes A^\dagger_{k_2}A_{k_3}\otimes
\cdots\otimes A^\dagger_{k_p}A_{k_1} 
   \ee
 was used.
However, \eqref{theta}  satisfies (\ref{linearize})
only when the input is a pure state.
In \cite{GLR},   the operator
\begin{align}
   \Omega(\Phi,p) \equiv \wh{\Phi}^{\otimes p}(L_p) 
\end{align}
 was introduced where  $ \wh{\Phi} $ denotes the adjoint with 
respect to the
Hilbert-Schmidt inner product and $L_p$ and $R_p$ are
   the left shift and the right shift operators
\begin{align*}
L_p\kt{k_1 k_2 \cdots k_p}&=\kt{k_2\cdots k_p k_1},\\
R_p\kt{k_1 \cdots k_{p-1} k_p}&=\kt{k_p k_1\cdots k_{p-1}},
\end{align*}
$ \Omega(\Phi,p)$ was shown to give a valid realization of $X$ for arbitrary 
$\rho$ and satisfy
\be
 \Omega(\Phi,p)  =  \Theta(\Phi,p) L_p
 \ee
We now give some relations between these operators and those
of their conjugates.

\begin{thm}\label{thm:G-L}
Let $\Phi$ be a CPT map and let $ \Theta(\Phi,p)$ and  $\Omega(\Phi,p)$
be the linearizing operators defined above using a fixed set of Kraus
operators.  Then
\be
   \Omega(\Phi,p)  ~ = ~ \Theta(\Phi^C,p)^\dagger  ~ = ~ 
\Theta(\Phi,p) L_p .\ee
\end{thm}
when $\Phi^C$ is defined used the Kraus representation given 
by \eqref{conj-Kr}.

\prf  The key point is that \eqref{conj-Kr} implies that conjugate sets of
 Kraus operators satisfy
\be
   \bra m |  F_\mu =  \sum_j  \bra j |  (F_\mu)_{mj}  
         =  \sum_j  \bra j |  (F_m)_{\mu j} =    \bra m |  R_m.
\ee
Then
   \begin{align}
   \Omega(\Phi,p)&= \wh{\Phi}^{\otimes p}(L_p)
   =\wh{\Phi}^{\otimes p}
     (\sum_{k_1,\cdots,k_p}
         \kt{k_2\cdots k_p k_1}
                      \braa{k_1 k_2 \cdots k_p})\nonumber\\
&= \sum_{k_1,\cdots,k_p}
   \wh{\Phi} 
(\kt{k_2}\braa{k_1}) \wh{\Phi}(\kt{k_3}\braa{k_2})\otimes
   \cdots\otimes
   \wh{\Phi} 
   (\kt{k_1}\braa{k_p})\nonumber\\
&= \sum_{k_1,\cdots,k_p}\sum_{\mu_1,\cdots,\mu_p}
    F^\dagger_{\mu_1}\kt{k_2}\braa{k_1}F_{\mu_1}\otimes
    F^\dagger_{\mu_2}\kt{k_3}\braa{k_2}F_{\mu_2}\otimes
    \cdots\otimes
    F^\dagger_{\mu_p}\kt{k_1}\braa{k_p}F_{\mu_p}
   \nonumber\\
&= \sum_{k_1,\cdots,k_p}\sum_{\mu_1,\cdots,\mu_p}
    R^{\dagger}_{k_2}\kt{\mu_1}\braa{\mu_1}R_{k_1}
     \otimes
     R^{\dagger}_{k_3}\kt{\mu_2}\braa{\mu_2}R_{k_2}
     \otimes  \cdots\otimes
    R^{\dagger}_{k_1}\kt{\mu_p}\braa{\mu_p}R_{k_p}\nonumber\\
&= \sum_{k_1,\cdots,k_p}
    R^{\dagger}_{k_2}R_{k_1}
     \otimes
     R^{\dagger}_{k_3}R_{k_2}
     \otimes  \cdots\otimes
    R^{\dagger}_{k_1}R_{k_p}.
\label{linear1}   \qed
\end{align}
Theorem~\ref{thm:G-L}  allows one to compute $\Omega(\Phi,p)$ from the Kraus
operators of $\Phi^C$ without using shift operators.    Conversely, one
can compute $\Theta(\Phi,p) = \Omega(\Phi^C,p)$ directly in terms of
the action of $\Phi^C$ on components of shift operators, without
knowing its Kraus expansion
or requiring a final multiplication by a shift operator.

\bigskip
\noindent {\bf Acknowledgements:}    The work of CK and MN
was supported in part by National Science Foundation Grant
DMS-0400426.     The work of KM was partially supported  by
the   ERATO Quantum Computation and Information Project
of the    Japan Science and Technology Agency.
The work of MBR and MN was partially supported  by
   by the National Science  Foundation under Grant  DMS-0314228 
and by  the National Security Agency  nd
 Advanced Research and Development Activity   under
Army Research Office   contract number    DAAD19-02-1-0065. 
   This work is an outgrowth of discussions between MBR and KM
during the program on Quantum Information at the Isaac  Newton Institute
at Cambridge University in 2004.   Some of the work of MBR was performed
during a visit to the ERATO Quantum Computation and Information Project
and National Institute of Informatics in Tokyo.
Some of the work of MN was performed at Northeastern University.

\bigskip

 
 \appendix

\section{Representations of CPT maps}  \label{app:rep}

We review here some facts about representations of CPT maps
on finite dimensional spaces.   For proofs and additional
details about the history we recommend Chapter 4 of Paulsen \cite{Paul}.

In more general situations, a CPT map is defined as the dual of a
unital CP map and some theorems are more conveniently stated
for unital maps.    In finite dimensions, a linear map 
$\Phi: M_d \mapsto M_{d^\prime}$ is trace-preserving if and only
if its dual $\wh{\Phi} :M_{d^\prime} \mapsto M_d$ is unital, where
\be  \label{dual}
    \tr  [\wh{\Phi}(A)]^\dag B = \tr A^\dag \Phi(B),
\ee
Here, we will state  results for unital maps in terms of $\wh{\Phi} $.

The first and most fundamental result is due to Stinespring \cite{Stine}.
 \begin{thm}  {\em (Stinespring)}
 Let $\Psi: {\cal A} \mapsto {\cal  B}({\cal K})$ be a CP map
 from the $C^*$-algebra ${\cal A}$  to the bounded operators 
 on the Hilbert space ${\cal K}$.  There exists a *-homomorphism
 $\pi: {\cal A} \mapsto {\cal  B}({\cal H})$ from ${\cal A}$  to the bounded 
 operators  on the Hilbert space ${\cal H}$ and a bounded operator
 $V: {\cal H} \mapsto {\cal K}$ such that
 \be   \label{stinerep}
   \Psi:(A) = V^{\dag} \, \pi(A) \, V .
 \ee
 Moreover, $\Psi $ is unital if and only if $V^{\dag}  V = I$.
  \end{thm}
 
 This result may seem strange to those familiar 
with the operator sum representation; it has the same form,
but with only a single term.
 However, the sum is hidden in the representation which can
 contain multiple copies of ${\cal A} $.    In fact, for
$\Psi = \wh{\Phi} $ with $\Phi$ a CPT map as above,
 one can show  \cite{Paul} that  
 $\pi(A) = A \ot I_{\kappa} = \sum_k A \ot \proj{e_k}$ with $\kappa \leq d \d2$.
 Then defining $F_k = (I_d \ot \bra e_k |) V$, we can write
$V = \sum_k F_k \ot |e_k \ket$ and 
 \be     \label{StoK}
  \wh{\Phi}(A) =   V^\dag  A \ot I_{\kappa}  V
         = \sum_k F_k^{\dag} A F_k 
  \ee
  with $\sum_k F_k^{\dag}  F_k  = V^\dag V = I_{\d2}$.   This is equivalent to
  the usual Kraus-Choi operator sum representation since,
 for any $A \in C^\d2, B \in C_d$,
\be
   \tr A^\dag  \Phi(B) = \tr  \sum_k [F_k^{\dag} A F_k ]^\dag B = 
        \tr A^\dag \Big( \sum_k F_k B F_k^\dag \Big),
\ee
  which implies $\Phi(B) = \sum_k F_k B F_k^\dag$.
  
  Moreover, 
  \be
     \tr A^\dag  \Phi(B) & = &  \tr (A \ot I_\kappa)^\dag \, V B V^\dag \nn \\
      & = &   \tr (A \ot I_\kappa)^\dag \, \Big( \sum_{jk} F_j B F_k^\dag  \ot |e_j\kb e_k|  \Big) \nn  \\
     & = &   \tr   (A \ot I_\kappa)^\dag \, U B \ot \proj{e_1} U^\dag 
  \ee
    where $U =  \sum_{jk} U_{jk}  \ot |e_j\kb e_k| $ with each $U_{jk}$ a $d \times \d2$
    matrix and  $U_{j1} = F_j$ or, equivalently, the first $\d2 $ columns of $U$ equal
    $V$.   Since $V^\dag V = I$, this implies that $U$ can be chosen to be a partial
    isometry of rank $d \kappa$ so that when $d = \d2$, $U$ is a unitary extension of $V$.
    Thus we conclude that any CPT map  can be represented in the form
    \be   \label{lind}
      \Phi(B) = \trp_2 U B \ot \proj{e_1} U^\dag
    \ee 
    with $U$ a partial isometry.   This is sometimes referred to as the ``Stinespring
    dilation theorem'', although \eqref{lind} does not appear explicitly in \cite{Stine}; 
Kretschmann and  Werner \cite{KW} use the termÊ``ancilla representation''.   As far as we are
    aware Lindblad \cite{Lind75} was the first to explicitly us a representation of the 
    form   \eqref{lind} and we will refer to it as the Lindblad-Stinespring ancilla
    representation.

Next, we consider the Choi-Jamiolkowsi (CJ) representation of a CP map 
\be   \label{CJ}
   \Gamma  =  (I \ot \Phi) ( \proj{\phi}) = 
      \frac{1}{d} \sum_{jk} |e_j \kb e_k| \ot  \Phi( |e_j \kb e_k|)
\ee
where $|e_i\ket$ denotes the standard basis for ${\bf C}^d$ and 
$|\phi \ket = d^{-1/2} \sum_{k} |e_k \ot e_k \ket$ a maximally entangled state.   
The condition that $\Phi$ is also trace-preserving becomes $\trp_A \, \Gamma =  \tfrac{1}{d}I_d$
Let   $z^\mu,  \mu = 1 \ldots \kappa$ denote the normalized eigenvectors of  $\Gamma$ with
 a non-zero eigenvalue.   Then
$     \Gamma  = \sum_{\mu = 1}^\kappa  \lambda_\mu |z^\mu \kb z^\mu | $.
Moreover,  the identification  
   $g^\mu_{mj} = \sqrt{ d \lambda_\mu} \, z^\mu_{(d^\prime -1 )m + j}$, gives
   a set of Kraus operators  $G^\mu$ for the channel, and
   \be
        \Gamma = \tfrac{1}{d} \sum_{jm,kt}  \sum_\mu g^\mu_{mj} \ovb{g}^\mu_{nk } 
             |e_j \ot e^\prime_m \kb e_k \ot e^\prime_n|.
   \ee
   where $|e^\prime_m\ket$ is the standard  orthonormal basis for $C_\d2$.
Note that $\kappa$ is the minimal number of Kraus operators and (up to degeneracy
of eigenvectors) this provides a canonical way of defining a set of Kraus operators
and shows that the minimal number is no greater than $d d^\prime$.   

\noindent{\bf Proof of Proposition \ref{CJconj}:}   
We can regard $\Gamma$  as a density matrix $\Gamma_{AB}$ on the tensor product space
${\bf C}_d \ot  {\bf C}_\d2$ and obtain a purification
\be
       \Gamma_{ABC} & = &   \sum_{\mu, nu} | \Psi_\mu \ot e^{\prime \prime}_mu \kb \Psi_L \ot e^{\prime \prime}_nu|  \nn \\
              & = &  \sum_{\mu, \nu} \sum_{mj,nk}    g^\mu_{mj} \ovb{g}^\nu_{nk } 
             |e_j \ot e^\prime_m  \ot e^{\prime \prime}_\mu \kb e_k \ot e^\prime_n \ot e^{\prime \prime}_L|
\ee
with  $\ds{ \Psi_\mu = \sum_{jm }  g^\mu_{mj}  |e_j \ot e^\prime_m  \ket }$ and 
Then, taking the partial trace over $\hil_B$ gives
\be
   \Gamma_{AC} =  \sum_{\mu,\nu} \sum_{jkm}    g^\mu_{mj} \ovb{g}^\nu_{ mk } 
             |e_j \ot  e^{\prime \prime}_\mu \kb e_k  \ot e^{\prime \prime}_L|
\ee
which has eigenvectors $\sum_{j \mu}  g^\mu_{mj}  |e_j \ot  e^{\prime \prime}_\mu \ket  $.
  Thus,  $f^m_{\mu j} = g^\mu_{mj} ,  m = 1 \ldots \d2$
form a set of Kraus operators for $\Phi^C$ and  $ \Gamma_{AC}$ is the CJ matrix
$(I \ot \Phi^C)( \proj{\phi}) $.   \qquad \qed

It is  well-known  (see \cite{Paul}, Proposition~4.2) that any two minimal representations
in Stinespring's dilation theorem are unitarily equivalent.    Indeed, this is the reason
there is no loss of generality in assuming that $\pi(A) = A \ot I_\kappa$ in \eqref{StoK}.
Similarly, it is easy to show that any two minimal sets of Kraus operators  are related
by a unitary transformation.   However, it is often useful to consider non-minimal sets,
in which case, the unitary transformation may be replaced by a partial isometry.
Since this situation may be less familiar, we make a precise statement.
\begin{thm}
  If $\{ G_k \}$ is a minimal set of Kraus operators for 
the CPT  map $\Phi$  and $U$ is partial isometry with $U^\dag U = I_\kappa$,
then 
\be    \label{K2}
F_j = \sum_k  u_{jk} G_k
\ee
 is also a set of Kraus operators.   Moreover, 
any two sets of Kraus operators $\{ F_j \}$ and $\{ F^\prime_j \}$ define the same
CPT map if and only if one can find a partial isometry $W$ of rank $\kappa$ such
that  $F_j = \sum_k  w_{jk} F^\prime_k$.   
\end{thm}
\prf  The first assertion is easy to verify.    Moreover, any set of Kraus operators
defines a set of vector of length $d \d2$  whose span is the range of the CJ matrix.
When one set $G_K$ is minimal, as in \eqref{K2}, the requirement that $\Phi$ is
trace-preserving implies that $U^\dag U = I_\kappa$.   
When both
$\{ F_j \}$ and $\{ F^\prime_j \}$ they must satisfy \eqref{K2} 
  with $G_k$ minimal and   $U, U^\prime $  partial isometries of rank $\kappa$.
Then  $G_k = \sum_j \ovb{u}_{jk} F_j$.  Therefore, 
$F^\prime_j = \sum_k \sum_m u^\prime_{jk}  \ovb{u}_{mk} F_m$ and $W = U^\prime U^\dag$
satisfies $ W W^\dag = U^\prime U^\dag U (U^\prime)^\dag = U^\prime  (U^\prime)^\dag $,
which is a projection of rank $\kappa$.     Although, we do not have $W^\dag W = I$,
reversing the roles of  $\{ F_j \}$ and $\{ F^\prime_j \}$ gives 
$F_j^\prime = \sum_k  v_{jk} F_k$ with $V = U (U^\prime)^\dag = W^\dag$.

\section{Qubit channels}  \label{app:qbit}

 In the case of
qubits, the decomposable images have the form
$\half N^C\big[I \pm \sigma_j]$ with $j = 1,2,3$ and are
permutationally equivalent to
$\tfrac{1}{4} \pmx   1 & 1 & 0 & 0 \\   1 & 1 & 0 & 0 \\ 0 & 0 & 1 & \pm i \\
         0 & 0 & \mp i & 1\emx $.     

 A   channel of the form \eqref{Cort} can be rewritten as
$\Phi(I + \bw \cdot \sigma) = I + \sum_k \lambda_k w_k \sigma_k$ with
$\half(1 + \lambda_k) = a_0 + a_k$ and
$\half(1 - \lambda_k) = a_i + a_j$ (with $i,j,k$ distinct).  With $k^*$ chosen
so that $| \lambda_{k^*}| \geq |\lambda_j| ~~ \forall ~ j = 1,2,3$, one finds
\be
     \nu_p(\Phi) & = &   \Big( \big[\half(1 + \lambda_{k^*})\big]^p +
\big[\half(1 - \lambda_{k^*})\big]^p \Big)^{1/p}     \nn \\
             & = &   \big[ (a_0 + a_{k^*})^p +  (a_i + a_j)^p \big]^{1/p}
     \ee
where we used the convention that  $i,j, k^*$ are distinct.    When
$\lambda_{k^*} > 0$,
$a_0$ and $a_{k^*}$ are the two largest coefficients;  when
$\lambda_{k^*} <  0$, they
are the two smallest.   Thus, the bound \eqref{upbd} is  attained
with either of
the two decomposable matrices  $\half N^C\!(I \pm \sigma_{k^*})$.
 
 In general the upper bound  \eqref{upbd}  is   {\em not}  attained for the
 product $\Phi \ot \Phi$.       To simplify the discussion, we now assume that
 $a_0 > a_1 > a_2 \geq a_3$ which implies 
 $1 > \lambda_1 >  \lambda_2 \geq   \lambda_3 $ and $\lambda_2 > 0$
 and involves no fundamental loss of generality.   (One can always
 conjugate with $\sigma_k*$ to make $a_0$ the largest and rotate
 axes to make $a_1 $ the second largest.)   King \cite{King2} showed
 that all unital qubit channels are multiplicative for all $p \geq 1$.    Therefore
the eigenvalues of the optimal output of $ \Phi \ot \Phi$ are
$\beta_1^2,  ~ \beta_1 \beta_2 ,  ~ \beta_2 \beta_1, ~ \beta_2^2$ with 
$$\beta_1^2 = (a_0 + a_1)^2, \quad  \beta_1 \beta_2 = (a_0 + a_1)(a_2 + a_3) =
 \beta_2 \beta_1 , \quad  \beta_2^2 = (a_2 + a_3)^2.$$
The first term in the upper bound  equals $\beta_1^2$, if and only if
 $a_1^2 \geq a_0 a_2$.  Then the ordering of the product coefficients 
 begins
\bee
    a_0^2 > a_0 a_1 = a_1 a_0 > a_1^2 ~~  >  ~~  a_0 a_2 = a_2 a_0 \ldots
\eee
so that the second term in the upper bound is either $2 a_0(a_2 + a_3)$
or $2(a_0+ a_1)a_2$, neither of which equals $ \beta_1 \beta_2$.  Thus,
the upper bound \eqref{upbd} is never attained with distinct $a_k$.
It is achieved if $a_0 = a_1$ and $a_2 = a_3$, but this is a QC channel.

Although the multiplicativity of unital qubit channels was established
in \cite{King2}, it would be desirable to prove this by the methods developed 
here.   This requires two additional assumptions
\begin{itemize}

\item[a)]  The state in ${\cal N}^\cT$ which achieves $\nu_p( \Phi \ot \Phi)$
for a unital qubit channel $\Phi$ is decomposable when $\cT = \cP \ot \cP$ is the
product Pauli basis.

\item[b)]  When $d = 4$ and $\cT$ is the product Pauli basis, 
all decomposable 
states  in ${\cal N}^{\cP \ot \cP}$ are either tensor products of axis states 
for $d  = 2$ or maximally
entangled states formed from evenly weighted superpositions of axis states.

\end{itemize}
Although (a) seems like a reasonable conjecture, we have no proof.   
Theorem~\ref{thm:prod-ME} implies (b); for qubits, an explicit
computation can show that these maximally entangled states must have
the form of the usual Bell states in one of the three axis bases.
  Then a direct comparison
of the short list of possible decomposable states shows that
$\norm{ \gamma A \gamma }_p$ is always less for the maximally
entangled states than for the optimal product.    This is a
tedious process which would be impractical even if  (a) holds
inÊ higher dimensions,   Nevertheless, it gives some insight and 
is reminiscent of the argument used in \cite{KR1}.

The next example exploits the isomorphism ${\bf C}_4 \simeq {\bf C}_2 \ot {\bf C}_2$
to show that decomposability
is a basis dependent property.   
\begin{exam} {\em If $\rho = \proj{v} $ with $\bra v | = (1, 0, 1, 0)$  then
$N^C(\rho)$ is decomposable in the generalized Pauli basis for $d = 4$, but
{\em not} is the basis given by products of (the usual) Pauli matrices.
  If $\rho = \proj{v} $ with $\bra v | = (1, 0, 0, 1)$  then
$N^C(\rho)$ is {\em not} decomposable in the generalized Pauli basis for $d = 4$, but
{\em is } decomposable in the basis given by products of (the usual) Pauli matrices.
} \end{exam}

\end{document}